**Inverse designed resistive heaters for uniform switching of Phase Change Materials**

Virat Tara[1], Khoi P. Dao[2], Juejun Hu[2] and Arka Majumdar[1,3]

[1]Department of Electrical and Computer Engineering, University of Washington, Seattle, WA, USA

[2]Department of Materials Science and Engineering, Massachusetts Institute of Technology, MA, USA

[3]Department of Physics, University of Washington, Seattle, WA, USA

Email: vtara@uw.edu, arka@uw.edu

## Abstract

Non-volatile phase-change material (PCM)-integrated metasurfaces offer a promising pathway toward next-generation solid-state reconfigurable free-space optics. However, their practical operation is currently bottlenecked by the highly non-uniform thermal profiles generated by the external heaters used to switch the PCM between its amorphous and crystalline states. The non-uniform heat profile in turn severely restricts the active switching area of the PCM integrated devices. In this work, we present an inverse-designed doped silicon resistive heater on a silicon-on-sapphire platform, featuring a spatially varying doping profile explicitly tailored to generate uniform heat. The new heater significantly improves spatial temperature uniformity, reducing the thermal gradient from 110 K in the case of typical conventional heater to merely 25 K in the case of the inverse-designed heater at a target temperature of ~1000 K. By meticulously optimizing the doping and annealing processes to suppress lateral dopant diffusion, we achieve a near-perfect spatial doping resolution capable of patterning two distinct doped Silicon filaments just 100 nm apart. We experimentally fabricate these devices and demonstrate their efficacy by successfully switching a large area ($\sim 18 \times 14\ \mu m^2$) of the wide-bandgap phase-change material (PCM) $Sb_2S_3$ using a compact heater geometry of size $26 \times 26\ \mu m^2$. We show that our inverse-designed heater achieves a 10-fold increase in active PCM switching area despite a reduction in the total heater footprint when compared to a conventional heater. Ultimately, this work provides a crucial steppingstone toward the development of non-volatile PCM-based reconfigurable free-space optics.

## Introduction

Phase-change material (PCM)-integrated metasurfaces[1][2][3][4] offer a promising pathway toward next-generation, non-volatile reconfigurable free-space optics that feature zero static power consumption, effectively eliminating the need for a continuous power supply to maintain a programmed state. As a modulation medium, PCMs are highly attractive due to their massive refractive index ($\Delta n \sim 1$) contrast[3] compared to conventional tuning mechanisms like thermo-optics, electro-optics, and liquid crystals. This means they can perform the same operation in a smaller form factor without any thermal or electric-field crosstalk in the steady state. The advent of wide-bandgap PCMs, such as GSST[5][6][7], $Sb_2S_3$[2][8][9], and $Sb_2Se_3$[10], has further advanced the field by enabling phase modulation alongside amplitude modulation. Switching PCMs between their amorphous (a-) and crystalline (c-) states requires localized thermal stimuli, which in photonics is predominantly supplied by doped silicon (Si) micro-heaters. The broad adaptability of doped Si heaters is evidenced by successful PCM switching demonstrations across the visible[11] to infrared (IR)[12] spectrum. Si is highly favored for this role due to its semiconductor nature, high refractive index which helps in tightly confining the optical mode, its transparency from the near-IR to longer wavelengths, and full CMOS compatibility.

Despite the widespread use of Si heaters, their fundamental design has largely remained unchanged, typically relying on uniformly doped active regions. While uniform doping is straightforward to implement, it inherently produces highly non-uniform thermal profiles[7][8][11]. This thermal gradient presents a major bottleneck for PCM-tunable metasurfaces because it strictly limits the effective switching area. To compensate and achieve uniform heating over a desired footprint, the overall heater size must be scaled up, which drastically reduces the pixel fill fraction in practical Spatial Light Modulator (SLM) architectures consisting of such PCM tunable pixels. To address this limitation, a theoretical inverse-design strategy was recently proposed[13]. By spatially varying the resistivity profile through precisely patterned, nanoscale doped Si filaments, this approach outlined a method to generate highly uniform heat distribution.

In this work, we present the first experimental realization of these inverse-designed heaters on a silicon-on-sapphire platform, achieving a thermal profile with a spatial temperature gradient of 25 K, compared to the 110 K gradient across a conventional heater at ~1000 K target temperature. To experimentally realize this design, we fabricated nanoscale doped Si filaments separated by gaps as narrow as 100 nm. This required overcoming the significant fabrication challenge of adjacent filaments electrically shorting due to lateral dopant diffusion. We mitigated this risk by developing an optimized shallow-doping procedure that employs heavy dopant species and precisely tuned rapid thermal annealing (RTA) to suppress lateral diffusion. Utilizing this refined fabrication process, we make the inverse-designed heaters and validate their superior thermal uniformity through infrared radiation imaging. Finally, we establish the practical efficacy of our device by demonstrating the reversible, large-area phase switching of the wide-bandgap PCM $Sb_2S_3$. We achieve a 10-fold increase in active PCM switching area when compared to a conventional heater. We believe this new class of inverse-designed doped Si heaters would become an integral part of future PCM based reconfigurable free-space optics.

## Results

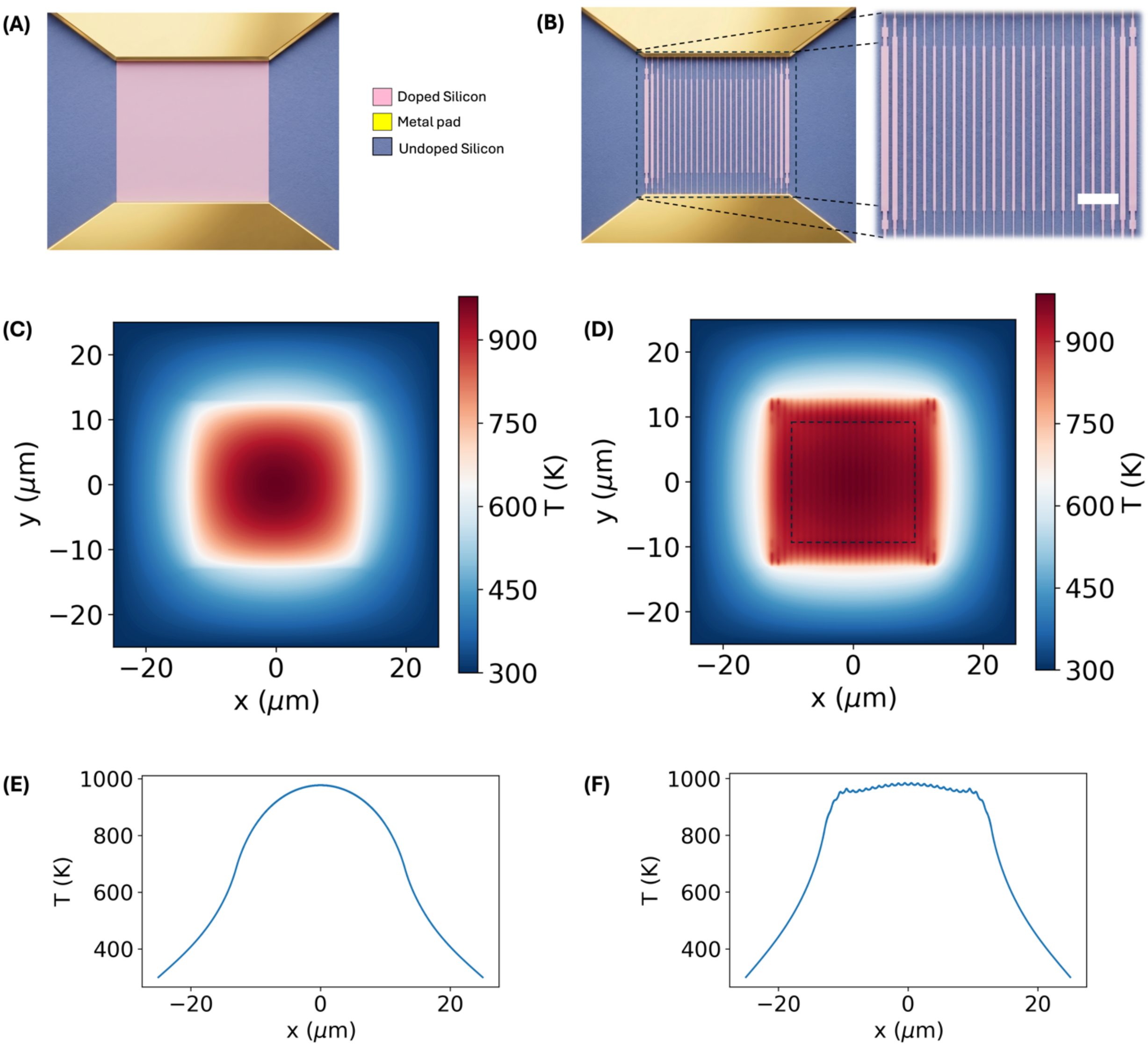

Figure 1. Device schematic and simulation (A) Schematic of a conventional uniformly doped Si heater. (B) Schematic of the inverse-designed Si heater with zoomed-in top view of the several conductive filaments that constitute one heater (scale: 4 μm). (C) Simulated heat profile of conventional and (C) Inverse-designed heater. The area inside the dotted square represents the target zone. (E, F) Temperature slice along y = 0 for conventional and inverse-designed heater respectively.

Figure 1 illustrates the device schematic and simulation results comparing the inverse-designed silicon (Si) heater with a conventional uniformly doped Si heater. The inverse-designed heater was developed following the procedure outlined in[13], with further details provided in the Methods section. Figures 1(A) and 1(B) illustrate the schematics of the conventional and inverse-designed heaters, respectively. The primary distinction between the two devices is the spatially varying resistivity profile of the inverse-designed heater, which is achieved by modulating the width of the conductive doped Si filaments connecting the two metal pads as shown in the zoomed-in view in Fig. 1(B). This variation in filament width locally alters the electrical resistance, thereby tailoring the spatial Joule heating distribution according to the relationship $P = \frac{V^2}{R}$, where $V$ is the voltage drop across a filament segment and $R$ is the electrical resistance of the segment. Because intrinsic Si is effectively non-conductive, these conductive pathways can be precisely defined through localized doping. The precise dimensions of the conductive filaments are given in the Supplementary Information. The inverse-designed heater spans a total area of $26 \times 26\ \mu m^2$, with a target zone[13] of $18 \times 18\ \mu m^2$ across which the thermal profile is designed to be uniform.

Thermal simulations of the conventional and inverse-designed heaters covering an area of $26 \times 26\ \mu m^2$ were performed using the Lumerical Charge and Heat solvers, with further details provided in the Methods section. As observed in the top-view thermal profiles, heat generation in the uniformly doped conventional device is primarily concentrated in the central region (Fig. 1(C)). This thermal non-uniformity arises because of the closer proximity of the outer edges to the surrounding Si film, which is at lower temperature. The inverse-designed heater mitigates this effect by structuring lower-resistance (wider width) filaments, which generate more heat near the boundaries and higher-resistance (smaller width) filaments in the center. This tailored spatial resistivity equalizes power dissipation across the device, resulting in a highly uniform temperature distribution from the center to the edges as shown in Fig. 1(D). This thermal homogenization is quantitatively illustrated by the temperature cross-sections taken along the y = 0 axis, as presented in Fig. 1(E) and 1(F). Notably, the spatial temperature variation (ΔT) for the conventional device is 110 K (difference between the temperature at X = 0 and X = 9 μm), whereas it is significantly reduced to merely 25 K for the inverse-designed heater at a target temperature of ~1000 K. The target temperature is well above the melting point of $Sb_2S_3$ (823 K) and hence enough to induce amorphization in $Sb_2S_3$. Since, amorphization of PCMs require rapid melt-quench of the material, we also simulate the transient response of our device. We note that the inverse heater achieves temperatures above the melting point of $Sb_2S_3$ in 20 $\mu s$ with more details in Supplementary Information. Several past works have shown that $Sb_2S_3$ can amorphize in $\mu s$ time scale pulses[8] [11][14]. We also present the temperature profile of the inverse designed heater close to the crystallization temperature of $Sb_2S_3$ (598 K) in the Supplementary Information.

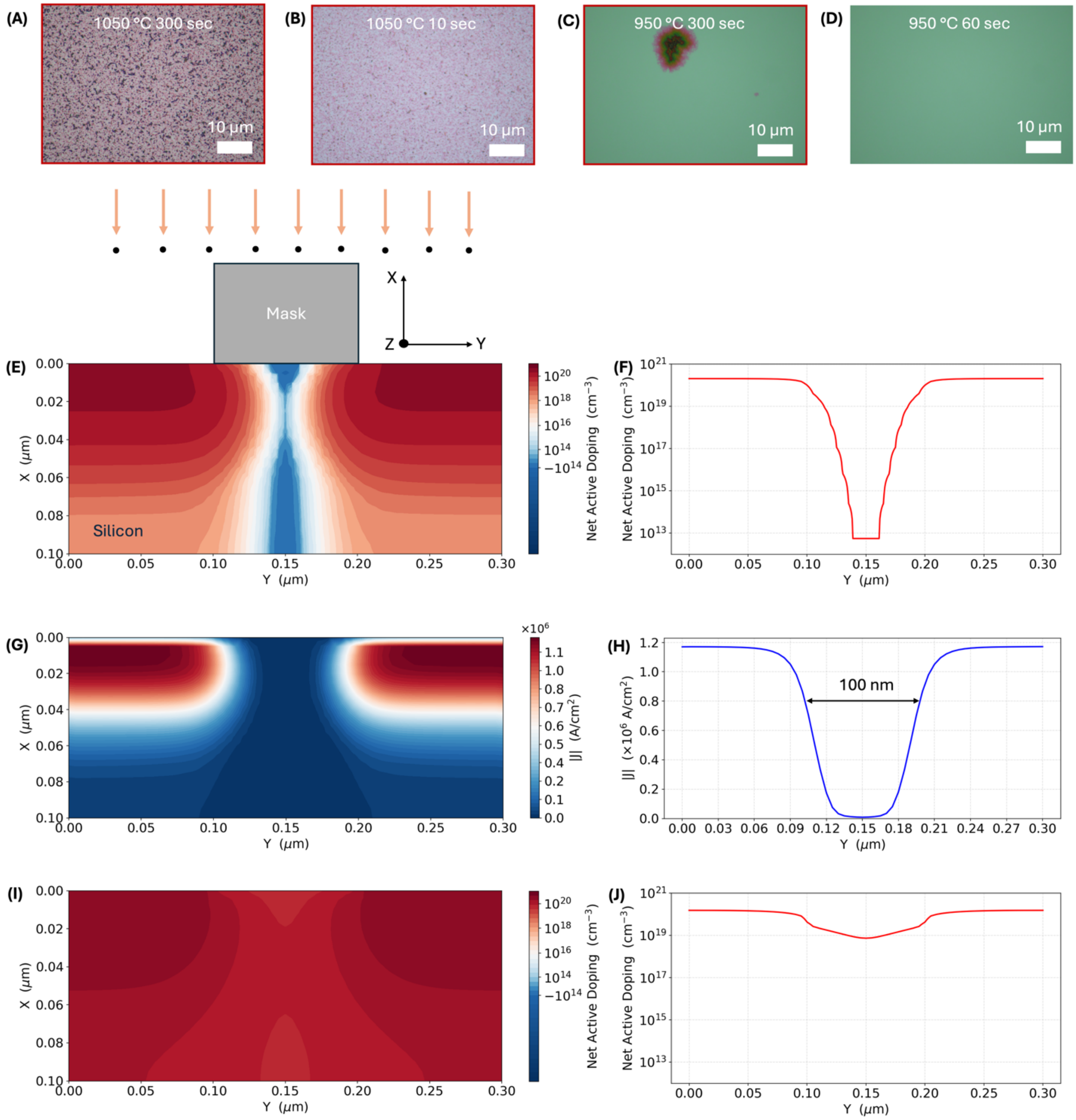

Figure 2. Experimental RTA tests and Doping simulation results (A, B, C) High temperature short duration annealing tests that resulted in damage to the Si layer. (D) 950 °C 60 second annealing that didn't induce any visible damage to the Si layer. (E) Simulated cross-sectional view of the doping concentration along the thickness of Si slab for Arsenic doping. Doping simulations were carried on a 100 nm thick (X) and 300 nm wide (Y) Si layer. A 100 nm wide $SiO_2$ mask was put on top of the Si layer to analyze how much the dopant diffuses laterally. (F) 1D slice plot of the doping concentration along X, Z = 0. (G) Cross-section view of the current density simulation showing the current density is primarily concentrated close the surface of Si and away from the $SiO_2$ mask. (H) 1D slice plot of the current density taken along X, Z = 0. (I, J) 2D and 1D doping

concentration plot for Phosphorus doping with the same parameters used for Arsenic doping simulation. Phosphorus being a lighter dopant compared to Arsenic diffuses faster.

Realizing the inverse-designed heater necessitates overcoming the fabrication challenge of defining ultra-thin (minimum 100 nm width) conductive filaments within the silicon layer. This means perfecting the two-step doping procedure: ion implantation, which is used to introduce dopants into the Si lattice, followed by thermal annealing to repair lattice damage and electrically activate the dopants via substitutional placement. A major constraint in this process is the characteristic lateral dopant diffusion. Therefore, to pattern these high-resolution conductive pathways on intrinsic silicon without causing adjacent filaments to electrically short, the dopant diffusion must be suppressed without adversely impacting dopant activation. Achieving this critical shallow-doping regime demands rigorous optimization of both the implantation and annealing parameters to minimize lateral diffusion and preserve the spatial resolution of the nanoscale filaments.

The initial ion implantation step can be optimized for shallow doping by employing either heavier dopant species, such as arsenic (As) or by utilizing steep sample tilt angles[15]. Due to their larger atomic mass, heavier dopants exhibit a shorter projected range during implantation and significantly reduced diffusivity compared to lighter dopants like boron (B) or phosphorus (P). This means, for a given implantation energy and thermal annealing time, a B or P atom will diffuse faster in the Si lattice than an As atom. Alternatively, while increasing the sample tilt angle yields shallower profiles, this approach typically is only suitable for uniform doping of a sample as using mask would also add a shadow to the doped area.

The second step, annealing, is critical because prolonged heating will diffuse even heavy dopants throughout the entire Si layer. To minimize dopant diffusion, we used rapid thermal annealing (RTA) for a duration of just a few seconds, which is sufficient to both repair the lattice damage (crystal regrowth) caused by ion implantation and electrically activate the dopants[16]. A major challenge with RTA, however, is that the rapid temperature fluctuations can easily damage the Si thin film on insulator substrate. To safely achieve shallow doping, we first tested various RTA recipes on silicon-on-insulator samples (Fig. 2(A–D)). We established that an RTA at 950 °C for 60 seconds (Fig. 2(D)) preserves the integrity of the Si layer. Since our RTA tool requires some stabilization time, the actual recipe used for RTA both in simulation and experiment is given in the Methods section.

Based on this optimized recipe, we then conducted dopant diffusion simulations using the Sentaurus Process module. We simulate a 300 nm wide (X), 100 nm thick (Y) and 300 nm long (Z) slab of Si with a 100 nm thick and wide $SiO_2$ mask. Figure 2(E) presents the cross-sectional doping profile of the simulated Si slab patterned with a 100-nm-wide $SiO_2$ mask following ion implantation and annealing. We use this simulation to evaluate the lateral diffusion of As under the 950 °C, 60 second annealing conditions. Our results indicate that an implantation energy of 20 keV combined with a dose of 1×$10^{15}$ ions/$cm^2$ effectively minimizes lateral diffusion. While the implant energy could be reduced even further for the same dose to reduce lateral diffusion, doing so would substantially increase the implantation costs as it takes longer to deliver the same dose at lower energies. Finally, the doping concentration slice taken along the x-axis at Z = 0 (Fig. 2(F)) clearly demonstrates that the doping level drops exponentially towards the center of the mask.

Despite some lateral diffusion of dopants as seen in Fig. 2(E), because current follows the path of least resistance, current density remains minimal in the moderately doped regions caused by lateral dopant leakage. To validate this behavior, we simulated the current density across the doped Si slab using Sentaurus Device module. Figures 2(G) and 2(H) demonstrate that the current is tightly confined to the heavily doped areas, resulting in negligible lateral leakage. Through these process optimizations, we successfully restrict the lateral diffusion of dopants. Finally, we contrast these results with the diffusion profile of phosphorus (P) atoms under the identical 950 °C, 60 second annealing conditions. As shown in Figures 2(I) and 2(J), the lower atomic mass of P causes rapid diffusion in both the vertical and lateral directions, which effectively short-circuits adjacent conductive filaments in our resistive heater. In the Supplementary Information we further show that by using even heavier dopant like Antimony (Sb)[17] we can precisely define the doped region depth, which potentially enables new photonic structures as proposed in [18] where the overlap between the optical mode and doped-Si is intentionally reduced to minimize absorption losses arising from doped-Si.

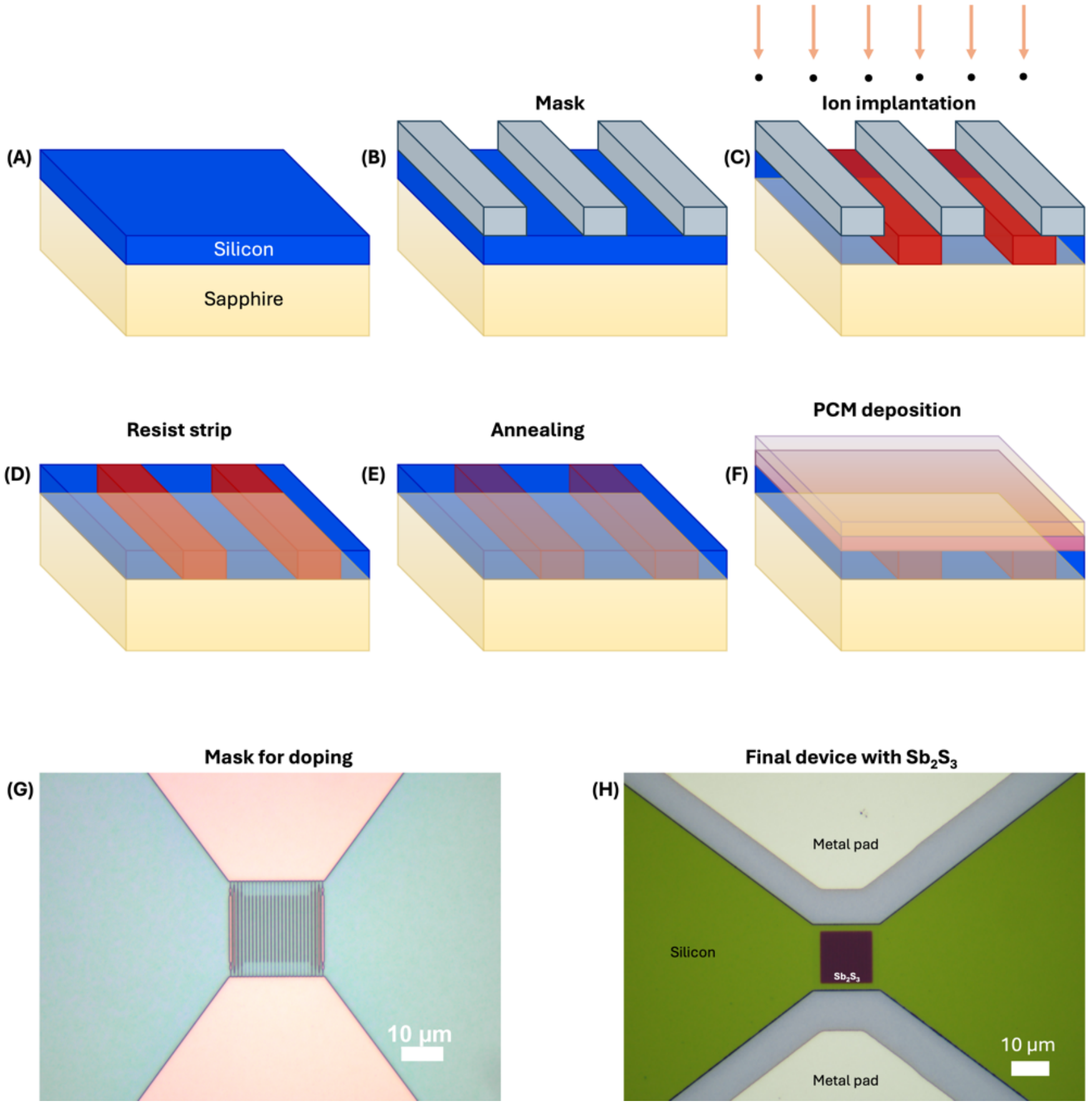

Figure 3. Fabrication of the inverse-designed heater (A) Starting 100 nm Silicon on Sapphire sample. (B) E-beam resist patterned to create a mask for ion-implantation. (C) Illustration of Arsenic ion-implantation. (D) Stripping of E-beam resist using Acetone. (E) RTA to diffuse and activate the dopants. (F) Illustration of PCM deposition and capping. (G) Micrograph of the device after E-beam resist patterning. (H) Micrograph of the final fabricated device.

Following the simulations, we fabricated the device according to the process flow illustrated in Figures 3(A–F). Fabrication begins with a 100-nm-thick silicon-on-sapphire substrate. Alignment marks are first etched into the sample to ensure precise alignment for all subsequent lithography steps. An ion-implantation mask is then patterned using a ~200 nm thick layer of ZEP1 (diluted ZEP0) electron-beam resist. Arsenic ions are implanted at an energy of 20 keV with a dose of $10^{15}$ ions/cm$^2$ and 7° tilt angle to avoid haphazard channeling. The resist is subsequently stripped with Acetone, and the sample undergoes rapid thermal annealing (RTA) at 950 °C for 60 seconds. Following dopant activation, 40 nm thick wide-bandgap PCM $Sb_2S_3$ is sputtered and protected with an 80 nm thick $Al_2O_3$ capping layer deposited via Atomic Layer Deposition (ALD). Procedures for fabricating the metal electrical contact pads are also detailed in the Methods section. Finally, Fig. 3(G) presents a micrograph of the device after resist development, while Fig. 3(H) displays the fully fabricated device with the $Sb_2S_3$ layer positioned atop the heater. More micrographs of fabricated devices and devices after ion-implantation are given in the Supplementary Information.

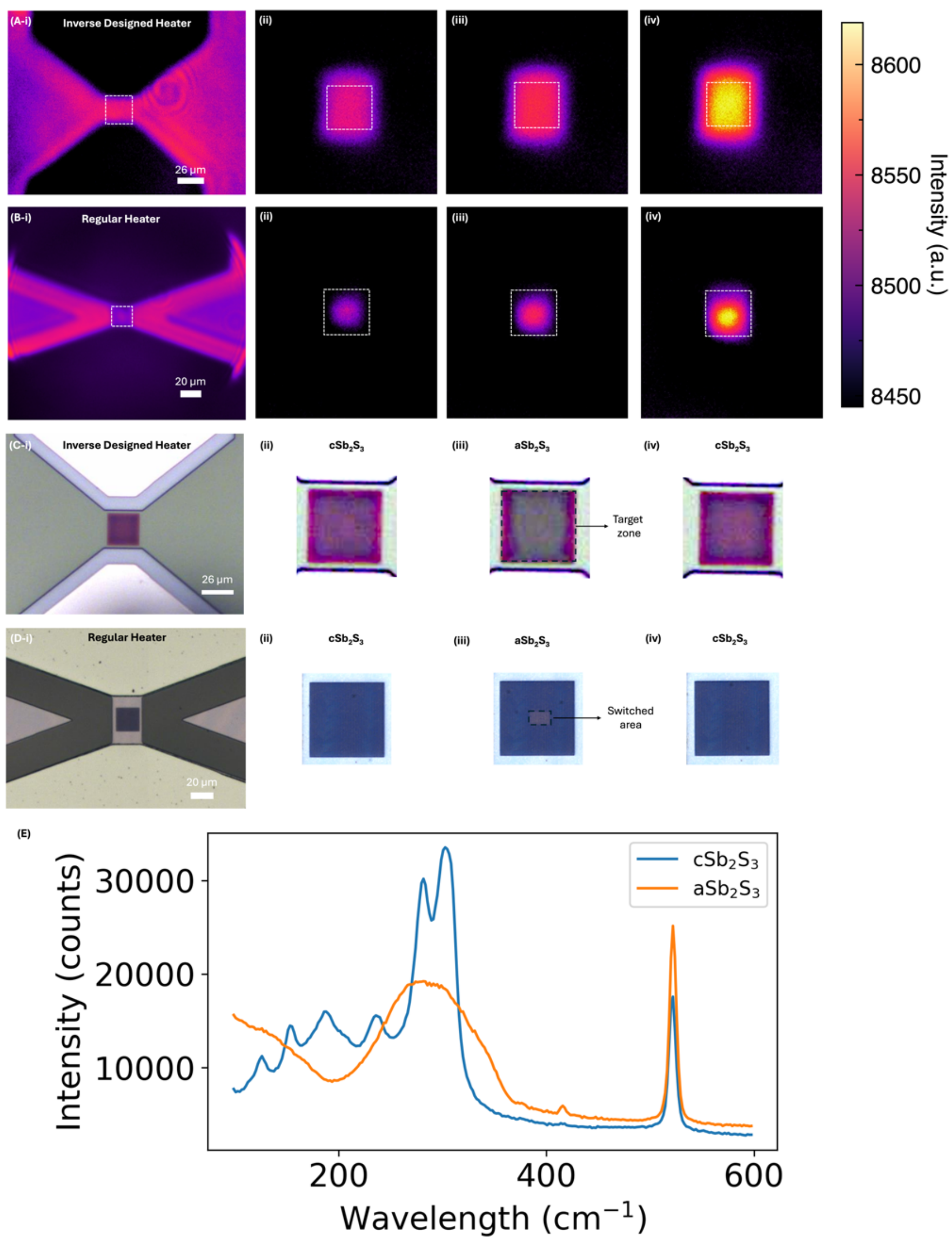

Figure 4. Measurement results (A-i) Device with $26 \times 26\ \mu m^2$ inverse designed heater imaged using an infrared camera. (A-ii, iii, iv) The thermal radiation measured from the heater at three different voltages with the applied voltage increasing from left-to-right. White dotted square denotes the heater area. (B-i) A device with $22 \times 30\ \mu m^2$ regular, uniformly doped Si heater imaged using the infrared camera. (B-ii, iii, iv) The thermal radiation measured from the heater at three different voltages applied in increasing order as we go from left-to-right. White dotted square denotes the heater area. (C-i) The final fabricated device with c-$Sb_2S_3$. (C-ii, iii, iv) Reversible switching of the PCM using the inverse designed heater. We can see the color change of the PCM

layer under the microscope. (D-i) A non-uniform heater device with c-$Sb_2S_3$. (D-ii, iii, iv) Reversible switching of the PCM using the regular non-uniform heater. We can again see the color change of the PCM under the microscope. Only a tiny portion of PCM is switched at the center of the heater. (E) Raman spectra of devices with electrically switched c-$Sb_2S_3$ and a-$Sb_2S_3$. The c-$Sb_2S_3$ Raman spectra show multiple peaks between 100 to 400 $cm^{-1}$ due to the crystalline nature of the material.

Following device fabrication, we characterized the samples, with the measurement results summarized in Figure 4. The measured resistance of one of the fabricated devices was ~1 kΩ. Future reductions in device resistance can be achieved by increasing the ion-implantation dose to $2\times10^{15}$ ions/$cm^2$ and extending the annealing time to 5 minutes. Doing this would essentially increase the doped Si thickness along the depth of the sample and reduce the device's sheet resistance. However, it would also increase lateral diffusion which can be readily pre-compensated in the lithography mask design. The IV curve of one of the fabricated devices is given in the Supplementary Information. The IV curve is perfectly linear indicating the presence of ohmic contacts. Due to the higher resistance of the device a high frequency power amplifier was used to apply signal to the inverse designed heater.

To visualize the spatial heat distribution, we imaged the thermal radiation using a mid-wave infrared (MWIR) camera. Figure 4(A-i) displays the device under external MWIR illumination. After switching off the external illumination, we applied three sequentially increasing voltages and imaged the resulting thermal emission (Fig. 4(A-ii–iv)). The voltages applied had 10 MHz frequency (square wave), 50% duty cycle and peak amplitude of 25.5V, 26.4V and 28.2V respectively. These images confirm the highly uniform heat generation of the inverse-designed heater, which becomes especially evident when contrasted with the localized thermal emission of a conventional heater (Fig. 4(B-i–iii)). The DC voltages applied to the regular heater were: 3.5V, 3.6V and 3.8V respectively. The dimensions of the regular heater used for IR imaging was $22 \times 30\ \mu m^2$.

Furthermore, we successfully demonstrated reversible phase-change switching of the $Sb_2S_3$ layer by applying voltage pulses (Fig. 4(C-i–iv)). Under optical microscopy, the transition between the crystalline (c-) and amorphous (a-) states is marked by a distinct color shift from red to green, which corresponds with the $Sb_2S_3$ bandgap change from 2.0 eV to 1.7 eV as it switches from a- to c- state. To induce amorphization we applied 1500 pulses at 10 MHz frequency with 50% duty cycle and 28 ns rise & fall time at peak voltage of 37V. To crystallize the PCM we applied a 10 MHz pulse train at 24.6V peak amplitude with 50 % duty cycle and 28 ns rise & fall time for several seconds. The black dotted square represents the target-zone. Roughly $18 \times 14\ \mu m^2$ PCM patch inside the target-zone of size $18 \times 18\ \mu m^2$ is successfully switched. Furthermore, we switched the device reversibly for 10 cycles. The micrographs of the device through these cycles are provided in the Supplementary Information.

In contrast, the conventional heater restricts PCM switching strictly to the central region. To induce amorphization we applied 30V 3.5μs pulse with a rise and fall time of 21 ns. To induce crystallization, we applied 12V 100ms pulse with a rise and fall time of 21 ns. The area switched in this case is just $\sim 5 \times 5\ \mu m^2$. We note that the switched area can be expanded by delivering higher power to the conventional heater. However, that would excessively heat the central part because of the large thermal gradient, risking irreversible damage to the PCM film. This device was extensively studied previously in[11] and has a dimensions of $25 \times 40\ \mu m^2$. Therefore, the inverse-designed heater successfully switches a 10-fold larger PCM area while simultaneously reducing the overall footprint of the device. Finally, we perform Raman spectroscopy on electrically crystallized and amorphized $Sb_2S_3$. Confirming the phase transitions corresponding to the change in color within the switched area on the inverse designed heater. The c-$Sb_2S_3$ spectrum exhibits multiple expected peaks[19] due to the crystalline nature of the material, whereas the a-$Sb_2S_3$ spectrum lacks these specific features, displaying only the prominent Si peak near 520 $cm^{-1}$.

## Conclusion

In conclusion, we have experimentally demonstrated an inverse-designed resistive heater capable of generating exceptionally uniform thermal profiles. By optimizing the implantation and annealing procedures, we achieved high-resolution replication of the nanoscale resistivity profile within the silicon substrate. The thermal performance of the fabricated devices was validated through infrared imaging and subsequently utilized to achieve large-area phase switching of the wide-bandgap PCM $Sb_2S_3$. Using the new heater design we can switch 10-fold more PCM area as compared to a conventional heater. The utility of these tailored thermal emitters extends broadly across reconfigurable photonics, well beyond PCM-tunable metasurfaces as the shallow doping used here can be leveraged in Si photonics modulators to minimize the overlap between the lossy doped region and the optical mode. Future work will explore the application of these inverse-design principles to achieve deterministic, multi-level PCM switching via intentionally engineered gradient heat profiles. Such advanced localized thermal control could unlock continuous-like phase modulation from 0 to $2\pi$ in tunable resonant metasurfaces using just one pair of electrodes per device. A further increase in the doping depth without causing excess lateral diffusion is required to lower the resistance of the device in the future. Additionally, the higher temperature gradient for the inverse designed heater in this work as compared to the simulation in ref. [13] can be attributed to two factors: (1) the design of the heater is optimized for 700 K temperature and (2) for Silicon with $SiO_2$ buried layer. In the future, the inverse design will be optimized directly for a silicon-on-sapphire platform, rather than an $SiO_2$ buried layer, to further minimize this thermal gradient. Ultimately, we anticipate that this new class of Si heaters will become a fundamental component for PCM based reconfigurable photonics.

**Methods**

*Simulation:* The optimization procedure followed the inverse-design approach described in ref. [13]. The heater unit cell was defined as a 1 × 1 $\mu m^2$ square tile consisting of a 100-nm-thick Si slab on a $SiO_2$ buried oxide layer. A series of localized simulations using COMSOL Multiphysics were performed to extract the thermal dissipation parameters of each unit cell. The inverse-design optimization was then carried out to achieve a uniform target temperature of 700 K. Post-inverse design, thermal simulations were performed using Lumerical Charge and Heat modules. The inverse-designed doping profile was imported onto a 100 nm thick Si on Sapphire substrate. The doping concentration of the Si layer was set as $10^{20}$ $cm^{-3}$. Since the inverse design heater uses shallow doping the thickness of doped Si in this case was kept as 40 nm. The shallow doping does result in an increase in the device resistance. The voltage applied to achieve ~1000 K in both conventional and inverse heater case was 3.25V and 12V respectively. To compensate the changes in heat distribution arising from higher thermal conductivity of sapphire as compared to $SiO_2$, the doped Si filaments width was further optimized to achieve uniform heat distribution. The final dimensions of the doped Si filaments used in this work are given in the Supplementary Information.

*Fabrication:* We start with 100 nm thick Si on Sapphire sample. We first pattern alignment marks onto the clean sample by spin coating ZEP1 resist and exposing it with E-beam lithography (JEOL 6300FS). After development alignment marks are etched using Reactive Ion Etching (RIE) with Fluorine gas (Oxford PlasmaLab 100). Following this we pattern the inverse designed heater filaments again using ZEP1 resist and E-beam lithography. Post development the sample was shipped for Arsenic ion-implantation. The parameters used for ion-implantation were 20 keV and implant dose of $10^{15}$ ions/$cm^2$ with a 7° tilt angle to avoid haphazard channeling. After ion-implantation the resist was stripped using 30 min sonication in Acetone and annealing was performed in RTA (Allwin 32 AccuThermo AW 610) at 950 °C for 60 seconds in $N_2$ environment. Since the tool requires some stabilization time the exact recipe used follows the below given trend:

| S. No. | Start temperature (°C) | Ramp rate (°C/s) | Annealing time (s) |
|---|---|---|---|
| 1. | 25 | 20 | 15 |
| 2. | 325 | 66.1 | 9 |
| 3. | 920 | 3 | 10 |
| 4. | 950 | 0 | 60 |
| 5. | 950 | -7.7 | 120 |

Post annealing, we again pattern the sample to put metal pads by spin coating PMMA-950K A6 and exposing it using E-beam lithography. 15 nm Titanium adhesion and 180 nm Platinum layers were deposited using E-beam evaporation (CHA SEC-600). The excess metal was lifted off by gentle sonication and leaving the sample in Acetone overnight. Next, the sample was again patterned by spin coating PMMA-495K A6 and exposed using E-beam lithography. Following which 40 nm $Sb_2S_3$ was sputtered on the developed sample. Immediately after $Sb_2S_3$ sputtering the PCM was capped using 80 nm $Al_2O_3$ grown via Atomic Layer Deposition (ALD) (Picosun

R200). $Al_2O_3$ was removed from the top of the metal pads by etching it using RIE with Chlorine gas (Oxford PlasmaLab 100). Lastly, the $Sb_2S_3$ was crystallized before performing electrical switching by annealing in RTA at 325 °C for 10 minutes.

*Experiment:* The thermal radiation measurements were recorded using a Mid-wave Infrared (MWIR) camera (FLIR A6751 MWIR InSb). The device with heater was imaged onto the camera sensor using a 2-inch focal length lens and two relay lenses. The Electrical signal was applied using probe stationers (Cascade Microtech DPP105-M-AI-S). The electrical pulses for measuring thermal radiation were generated by a pulse function arbitrary generator (Keysight 81160A) and amplified using a high-power amplifier LZY-22+ (for inverse designed heater). For the conventional heater we used a DC power supply (GW Instek GPE-2323). PCM switching for the inverse designed heater was carried out under a microscope using probe stationers and arbitrary function generator attached to the LZY-22+ amplifier. PCM switching for the regular heater was carried out using the arbitrary function generator and Cleverscope CS1070 amplifier with a 1-ohm output impedance. Raman spectroscopy was carried out using Renishaw InVia spectrometer.

## Acknowledgment

We thank Fred Newman for help with the RTA recipe development. This work is supported by the NSF-FuSe and NASA-STTR programs. Part of this work was conducted at the Washington Nanofabrication Facility / Molecular Analysis Facility, a National Nanotechnology Coordinated Infrastructure (NNCI) site at the University of Washington with partial support from the National Science Foundation via awards NNCI-1542101 and NNCI-2025489.

## Author Contribution

AM and JJ conceived the project. KD designed the inverse designed heater. VT performed the doping, annealing and the final thermal simulations. VT performed the device fabrication and measurement. KD helped with the device measurement. AM and JJ supervised the progress of the project. VT wrote the manuscript with inputs from all the authors.